\documentclass{article}

\def\be{\begin{equation}}
\def\ee{\end{equation}}

\begin{document}

\title{Coupling parameters  and the  form of the potential via Noether
 symmetry}
\author{A.K. Sanyal$^*$ \and C. Rubano$^{**}$ \and E.
 Piedipalumbo$^{***}$}
\maketitle

\noindent

\begin{center}
$^*$Dept. of Physics, Jangipur College, Murshidabad, \noindent India -
 742213%
\\[0pt]
\noindent and \\[0pt]
\noindent Relativity and Cosmology Research Centre Dept. of Physics,
 Jadavpur University
 Kolkata - 700032, India \\[0pt] $^{**}$Dept. of Physical Sciences -
 Univ.
``Federico II'', Via Cinthia, Naples, Italy\\[0pt]
\noindent and\\[0pt]
\noindent I.N.F.N. Section of Naples\\[0pt]
$^{***}$ I.N.A.F. - Osservatorio Astronomico di Capodimonte, via
 Moiariello,
Naples, Italy
\end{center}

\noindent
\footnote[1]{
\noindent electronic-mail: rubano@na.infn.it, aks@juphys.ernet.in}

\begin{center}
\textbf{{\Large {Abstract}} }
\end{center}

\noindent
 We explore the conditions  for the existence  of Noether symmetries
 in the
dynamics of FRW metric, non minimally coupled with a scalar field, in
 the
most general situation, and with nonzero spatial curvature. When such
symmetries are present we find general exact solution for the Einstein

equations. We also show that non Noether  symmetries can be found.
 Finally,
we present an extension of the procedure to the  Kantowski- Sachs
 metric
which is particularly  interesting in the case of degenerate
Lagrangian.\medskip

\section{\textbf{Introduction}}

The importance of \ `Scalar Tensor Theory of Gravity' is that both the

gravitational and the cosmological constants result from a single
 scalar
field which is somehow coupled with the curvature scalar. Both these
so-called constants are time dependent, dynamical quantities in the
 theory.
Further, different unification schemes of fundamental interactions
 based on
supergravity and superstring theories lead to scalar tensor theory of
gravity in the weak energy limit \cite{g:s}. The theory is also
 supposed to
produce successful phase transition and help to solve the problems
regarding graceful exit and the density perturbations \cite{l:s}.

Brans-Dicke \cite{b:d}, being motivated to incorporate Mach's
 Principle in
general theory of relativity, were the first to present a simple form
 of the
scalar tensor theory of gravity by means of a constant coupling
 parameter $w$%
. However, the experimental  constraints on $w$, resulting from some
classical tests of gravitation \cite{w:l}, naturally requires that $w$

should be a function of the scalar field $\phi $, rather than a
constant. The theory was thereafter generalized by Nordtvedt and
Wagoner \cite{b:n} to entertain arbitrary self interaction of the
scalar field in addition to the dynamical coupling with gravity. The
importance of the theory was increased by the work of Mathiazhagan and

Johri \cite{m:j}, where they proposed a revised model of the
inflationary Universe under the framework of Brans-Dicke theory .A
different form of scalar tensor theory was proposed by Zee \cite{a:z}
incorporating the concept of spontaneous symmetry breaking. In that
theory, which is known as the \ `Induced Theory of Gravity', the form
of the coupling with the scalar field with gravity is chosen as
$\epsilon \phi ^{2}$ , where $\epsilon $ is a dimensionless constant.
Though it was suggested that the same symmetry breaking mechanism is
responsible for breaking a unified gauge theory into the strong, weak
and electromagnetic interactions, yet such theory never goes over
asymptotically to th
e standard FRW model. Rather, it is much better to
choose the non-minimally coupled theory in its standard form, which
again arises as the low energy effective action of different string
theories \cite{g:s}. In such theory, the coupling of the scalar field
with gravity is taken as $(1-\zeta \phi ^{2})$, where $\zeta $ is a
dimensionless coupling constant. It is apparent that if $\phi $ admits

a solution that dies out with the cosmological evolution, the theory
leads to the standard model with a minimally coupled scalar field.

All these Brans-Dicke, induced and non-minimally coupled theories of
gravity are special cases of \ \ `Scalar Tensor Theory of Gravity',
 which
can be cast in such a general form that it even includes the
Einstein-Hilbert action for a minimally coupled scalar field as a
 special
case. The most general form of such a scalar tensor theory of gravity
 is
\begin{equation}
A=\int d^{4}x\sqrt{-g}\left( f(\phi )R-\frac{w(\phi )}{\phi }\phi
,_{\mu }\phi ^{,}{}^{\mu }-V(\phi )\right).  \label{eq:action}
\end{equation}
This form is the most general one, since for $f(\phi )=\phi $, it
 reduces to
the Brans-Dicke form, for $f(\phi )=\epsilon \phi ^{2}$, it takes the
 form
of the induced theory of gravity, for $f(\phi )=1-\zeta \phi ^{2}$ and
 $%
\displaystyle\frac{w}{\phi }=\displaystyle\frac{1}{2}$, it is of the
 form of
standard non-minimally coupled scalar field theory, for $f(\phi
)=\displaystyle\frac{\phi
^{2}}{6}$ and $w(\phi )=\displaystyle\frac{\phi }{2}$, the conformally
 coupled theory
can be obtained. Finally, for $f(\phi )=1/2$ and
 $\displaystyle\frac{w(\phi
)}{\phi }=\frac{1}{2}$ , it reduces to the form of Einstein-Hilbert
 action
minimally coupled with a scalar field.

Essentially the form of the coupling parameters $f(\phi )$, $w(\phi )$

and the form of the potential $V(\phi )$ are not known a-priori and
can not be obtained from the field equations. A new approach was
initiated by de Ritis and coworkers (\cite{c1:r}, \cite{c:r} and the
references therein) to find the forms of these parameters by demanding

that the Lagrangian admits Noether symmetry. As well known, the
Noether theorem states that if there exists a vector field $X$, such
that the Lie derivative of a given Lagrangian vanishes, then $X$ is a
symmetry for the dynamics and generates a conserved current. The
scheme was later carried out by Modak, Kamilya and Biswas \cite{b:k}
to find the form of $w(\phi )$. However, until very recently, none
considered the most general form of the action (1) that we are talking

of to get solution of the Einstein field Eqns. In~\cite{f:s} Fay
studies the exsistence of symmetries for this action. Our results are
more general, as we find exact solutions when possible. Moreover, in
the following paragraphs we discuss the importance of considering such

a general form of the action
.

It was a general belief that all the dynamical symmetries of a
 Lagrangian
can be extracted by the application of Noether theorem. That it is not
 so
has only recently been shown in a couple of publications by Sanyal and
 Modak
\cite{a:c} and Sanyal \cite{a:p}, (for a general discussion see
 \cite{g:p}).
It has been observed in \cite{a:c} that in the Robertson-Walker metric
 the
Noether symmetry makes the Lagrangian density $\ $(\ref{eq:action})
degenerate, if one looks for $f(\phi )$ in closed form. Moreover, the
 forms
of $f(\phi )$ and $V(\phi )$ thus obtained do not satisfy the field
equations for $k={{\pm} }{1}$. This very strange situation that
 Noether
symmetry does not satisfy the field equations has never been
 experienced
before. However, if $k=0$, there is no such trouble. Field equations
 are
well satisfied, degeneracy only leads to a constraint that has been
 analyzed
and solutions to the field equations are found. The same work in the
Kantowski-Sachs metric again reveals that the Lagrangian has to be
degenerate, but here also the field equations are satisfied.
 Degeneracy has
it's usual feature that leads to a constraint, which can be analyzed,
 and
solutions are obtained. It is thus clear that the very strange
 situation
that we came across in the FRW metric for $k={{\pm} }{1}$ has nothing
 to do
with degeneracy, but the reason of such outcome is still not known and

should be studied in order to understand Noether symmetry better.
 Moreover
such degeneracy leads to the very unpleasant consequence of leading to

negative Newton's gravitational constant.

Surprisingly enough, it was possible to explore some other type of
 symmetry
in both the situations \cite{a:c} and \cite{a:p}, which has nothing to
 do
with Noether symmetry, and which does not make the Lagrangian to be
degenerate. Such a situation when the symmetries are hidden and can
 not be
explored by Noether theorem is rather new and demands that symmetries
 of a
system should be studied thoroughly.

In the following section, we find the field equations for action (1).
 We
shall also find a couple of equations from the field equations, one of

which is found to be elegant in finding the Noether conserved current.
 The
importantce of the other equation is to study other forms of symmetry,

which can not be found by the application of Noether theorem. In Sec.
 3, we
apply Noether theorem and find the solutions of the equations thus
produced. As already mentioned, it requires to make additional
 assumptions
to obtain explicit solutions of the equations, which we do in Sec.~4.
 In
this section we study the situation case by case, find the conserved
current, express the Lagrangian in terms of the cyclic co-ordinates
 and
find explicit solutions of the field equations. Sec. 5, is devoted to
explore other forms of symmetries. In Sec. 6, we produce corresponding

results in the homogeneous, but anisotropic cosmological model, taking

Kantowski-Sachs metric as our starting point. Concluding remarks are
 made
in Sec. 7.

\section{\textbf{Action and Field Equations}}

Our starting point, as already mentioned in the introduction , is the
most general form of the scalar tensor theory of gravity, given by the

action~(\ref{eq:action}). In \cite{c:r}, \cite{c1:r} it is shown that,

by a suitable redefinition of the scalar field, it is possible to
bring this action in the form of a general nonminimally coupled
action. Thus it may appear that this kind of generalization is
useless. The simplest way to justify it is by means of an example,
which will be discussed in Sec.4, Case 2.

In the Robertson-Walker metric the Ricci scalar is
 $R=6(\frac{\ddot{a}}{a}+%
\frac{\dot{a}^{2}}{a^{2}}+\frac{k}{a^{2}})$. In view of which the
 action
takes the following form
\begin{equation}
A=\int \left( -6a^{2}\dot{a}\dot{\phi}f^{\prime
 }-6fa\dot{a^{2}}+6kfa+\frac{w%
}{\phi }a^{3}\dot{\phi}^{2}-a^{3}V\right) dt  \label{eq:action2}
\end{equation}

It should be noted that the Lagrangian turns out to be degenerate if
 the
Hessian determinant $W=\Sigma \frac{\partial ^{2}{L}}{\partial
 {\dot{a}}%
\partial {\dot{\phi}}}=0$. For the action (\ref{eq:action2}) it is $%
W=-12a^{4}(3f^{\prime }{}^{2}+2\frac{wf}{\phi })$. Therefore the
 Lagrangian
under consideration is degenerate under the condition
\begin{equation}
3f^{\prime }{}^{2}+2\frac{wf}{\phi }=0.
\end{equation}

The field equations are
\begin{equation}
2\frac{\ddot{a}}{a}+\frac{f^{\prime }}{f}\ddot{\phi}+\frac{f^{\prime
 \prime }%
}{f}\dot{\phi ^{2}}+\frac{w\dot{\phi ^{2}}}{2f\phi
 }+2\frac{\dot{a}f^{\prime
}\dot{\phi}}{af}+\frac{\dot{a^{2}}}{a^{2}}+\frac{k}{a^{2}}-\frac{V}{2f
}=0
\end{equation}
\begin{equation}
\frac{\ddot{a}}{a}-\frac{w\ddot{\phi}}{3f^{\prime }\phi }+\left(
 \frac{w}{%
\phi ^{2}}-\frac{w^{\prime }}{\phi }\right) \frac{\dot{\phi ^{2}}}{%
6f^{\prime
 }}+\frac{\dot{a^{2}}}{a^{2}}-\frac{w\dot{a}\dot{\phi}}{f^{\prime
}a\phi }+\frac{k}{a^{2}}-\frac{V^{\prime }}{6f^{\prime }}=0
\end{equation}
\begin{equation}
\frac{\dot{a^{2}}}{a^{2}}-\frac{w}{6f}\frac{\dot{\phi ^{2}}}{\phi
 }+\frac{%
f^{\prime }\dot{a}}{fa}\dot{\phi}+\frac{k}{a^{2}}-\frac{V}{6f}=0,
\label{eq:first}
\end{equation}
where, dot stands for derivative with respect to time and prime
 represents
derivative with respect to $\phi $. Eqn. (\ref{eq:first}) is a first
integral, which is obtained by the so-called ``Energy function'' of
Lagrangian (2). As we are considering the case of pure scalar field,
without matter, we have to set $E_{L}=0$. The case with matter can be
treated also, but we will do this in the future. Interesting results
 are
given in~\cite{f:s}.

By simple  algebraic  manipulations, it is possible to recast these
equations in a form which allows to find more general symmetries,
 discussed
in Sec. 5.
\begin{eqnarray}
&&\sqrt{\left( 3f^{\prime }{}^{2}+\frac{2wf}{\phi }\right)
 }~~~\frac{d}{dt}
\left( {\sqrt{3f^{\prime }{}^{2}+\frac{2wf}{\phi
 }}a^{3}\dot{\phi}}\right)
+a^{3}f^{3}\left( \frac{V}{f^{2}}\right) ^{\prime }=0 
 \label{eq:newform}\\
&&\frac{\ddot{a}}{a}+\left( \frac{f^{\prime }}{f}+\frac{w}{3\phi
 f^{\prime
}}
\right) \ddot{\phi}+\left( \frac{f^{\prime \prime
 }}{f}+\frac{w}{2f\phi }-%
\frac{w}{6f^{\prime }\phi ^{2}}+\frac{w^{\prime }}{6f^{\prime }\phi
 }\right)
\dot{\phi ^{2}}+\\ \nonumber&&\left( 2\frac{f^{\prime
 }}{f}+\frac{w}{f^{\prime }\phi }%
\right) \frac{\dot{a}}{a}\dot{\phi}+\frac{V^{\prime }}{6f^{\prime
 }}-\frac{V%
}{2f}=0.  \label{eq:current}
\end{eqnarray}

\section{\textbf{Application of Noether theorem}}

As already mentioned in Sec. 1, Noether theorem states that, if there
exists a vector field $X$, for which the Lie derivative of a given
Lagrangian $L$ vanishes i.e. $\mathcal{L}_{X}L=0$, the Lagrangian
 admits a
Noether symmetry and thus yields a conserved current. In the
 Lagrangian
under consideration the configuration space is $M=(a,\phi )$ and the
corresponding tangent space is $TM=(a,\phi ,\dot{a},\dot{\phi})$.
 Hence the
generic infinitesimal generator of the Noether symmetry is
\begin{equation}
X=\alpha \frac{\partial }{\partial {a}}+\beta \frac{\partial
 }{\partial {%
\phi }}+\dot{\alpha}\frac{\partial }{\partial
 {\dot{a}}}+\dot{\beta}\frac{%
\partial }{\partial {\dot{\phi}}},
\end{equation}
where $\alpha $ and $\beta $ are both  functions of $a$ and $\phi $
 and
\begin{equation}
\dot{\alpha}\equiv\frac{\partial \alpha }{\partial
 a}\dot{a}+\frac{\partial
\alpha }{\partial \phi }\dot{\phi}\quad ;\quad
 \dot{\beta}\equiv\frac{\partial
\beta }{\partial a}\dot{a}+\frac{\partial \beta }{\partial \phi
 }\dot{\phi}.
\end{equation}

The Cartan one form is
\begin{equation}
\theta _{L}=\frac{\partial L}{\partial{\dot{a}}}da+\frac{\partial
 L}{\partial {\dot{\phi}}}d\phi .
\end{equation}

The constant  of motion  $Q=i_{X}\theta _{L}$ is given by
\begin{equation}
Q=\alpha (a,\phi )\frac{\partial L}{\partial {\dot{a}}}+\beta (a,\phi
)\frac{\partial L}{\partial {\dot{\phi}}}.
\end{equation}

If $X$ is found, it is then possible to find a change of variables $%
u(a,\phi) $ , $v(a,\phi) $, such that
\begin{equation}  \label{eq:trans}
i_X du = 1 \quad ; \quad i_X dv = 0 .
\end{equation}

When the Lagrangian is expressed in the new variables, $u$ turns out
 to be
cyclic. The conserved current assumes a very simple form and we obtain
 often
exact integration.

If we demand the existence of Noether symmetry $\mathcal{L}_{X}L=0$,
 we get
the following equations,
\begin{equation}
\alpha +2a\frac{\partial {\alpha }}{\partial {a}}+a^{2}\frac{\partial
 {\beta
}}{\partial {a}}\frac{f^{\prime }}{f}+a\beta \frac{f^{\prime }}{f}=0
\label{eq:i}
\end{equation}
\begin{equation}
3\alpha \phi -a\beta +2a\phi \frac{\partial \beta }{\partial \phi
 }+a\phi
\beta \frac{w^{\prime }}{w}-6\frac{\phi ^{2}f^{\prime
 }}{w}\frac{\partial
\alpha }{\partial \phi }=0
\end{equation}
\label{eq:ii}
\begin{equation}
\left( 2\alpha +a\frac{\partial {\alpha }}{\partial
 {a}}+a\frac{\partial
\beta }{\partial \phi }\right) f^{\prime }+af^{\prime \prime }\beta
 +2f\frac{%
\partial \alpha }{\partial \phi }-\frac{wa^{2}}{3\phi }\frac{\partial
 {\beta
}}{\partial {a}}=0  \label{eq:iii}
\end{equation}
\begin{equation}
6kf\left( \alpha +a\beta \frac{f^{\prime }}{f}\right) =a^{2}V\left(
 3\alpha +%
\frac{V^{\prime }}{V}a\beta \right)   \label{eq:iv}
\end{equation}

We have now to look for conditions on the integrability of this set of

equations. We obtain restrictions on the form of $f$, $w$ and $V$, but
 large
freedom of choice will be left, so that all the interesting cases are
captured. Due to the very complicate situation we limit to the case
 when $%
\alpha $ and $\beta $ are separable (and non null), i.e.
\begin{equation}
\alpha (a,\phi )=A_{1}(a)B_{1}(\phi ),~~~\beta (a,\phi
 )=A_{2}(a)B_{2}(\phi
).  \label{eq:fact}
\end{equation}

We are thus well aware of the possibility of loosing some solutions
 (see
Sec. 5). We decided also to consider only the case $k\neq 0$, which is
 not
well treated in the literature, with the exception of case 6 in Sec.
 5. With
these assumptions, we show in Appendix that the conditions for
 integration
are
\begin{eqnarray}
&&A_{1}=-\frac{c~l}{a}\quad;\quad A_{2}=\frac{l}{a^{2}}\quad;\quad
V=V_{0}f^{3}\quad;\quad\frac{ B_{1}^{\prime }}{B_{1}}f^{\prime
}=\frac{2w}{3\phi }\quad;\quad\\ \nonumber &&\frac{B_{2}f^{\prime
}}{B_{1}f}=c\quad;\quad3f^{\prime }{}^{2}+\frac{2w}{\phi
}f=n\frac{wf^{3}}{\phi },
\label{eq:netcond}
\end{eqnarray}
where, $c,l,n,V_{0}$ are all arbitrary constants.

\section{\textbf{Solutions under different assumptions}}

In this section, we make reasonable assumptions on the form of $f$ or
 $w$,
in order to find the solutions of Eqns. (\ref{eq:i}-\ref{eq:iv}), The
transformation of variables (\ref{eq:trans}) and the new form of the
Lagrangian, which turns out to be surprising simple. The conserved
 current
and the $E_{L}$ function give now first order differential equations,
 which
can be exactly solved, in principle.\medskip

{\Large \textbf{Case 1}}. \medskip Let us consider the a general
nonminimally coupled case (that is: $w=\frac{\phi }{2}$), like in
\cite{c:r}. We get in view, of equation (\ref{eq:netcond}),
\begin{equation}
3f^{\prime }{}^{2}+f=\frac{n}{2}f^{3}.
\end{equation}

This gives an elliptic integral, which can be solved for $f$ in closed
 form
only under the assumption $n=0$, for which the Lagrangian turns out to
 be
degenerate. The solution is $f=-(\phi -\phi _{0})^{2}/12$, which means
 that
the Newtonian gravitational constant turns out to be negative. More
strikingly, the solutions obtained do not satisfy the field equations,
 as
one can easily verify from Eqn.~(\ref{eq:newform}) . It is still not
 clear
how to interpret such a situation. It is definitely not due to the
degeneracy of the Lagrangian, since the similar situation in the
Kantowsky-Sachs metric \cite{a:p} is found to behave properly. This is
 the
situation we came acrooss in \cite{a:c} and so we leave this case at
 this
stage. \medskip

{\Large \textbf{Case 2}}. \medskip

Let us now consider  a general Brans-Dicke theory (that is $f=\phi $).

Under this condition and in view of
\ Eqn. (\ref{eq:netcond} ), we obtain the following solutions
\begin{equation}
V=V_{0}\phi ^{3},~~~w=\frac{3}{n\phi ^{2}-2},~~~~n\neq 0,
\end{equation}
together with
\begin{equation}
B_{1}=B_{0}\frac{\sqrt{n\phi ^{2}-2}}{\phi
 },~~~B_{2}=cB_{0}\sqrt{n\phi
^{2}-2},
\end{equation}
where, $B_{0}$ is a constant. Hence $\alpha $ and $\beta $ are
 obtained as,
\begin{equation}
\alpha =-C\frac{\sqrt{n\phi ^{2}-2}}{a\phi },~~~\beta
 =C\frac{\sqrt{n\phi
^{2}-2}}{a^{2}},
\end{equation}
where $C=c~l~B_{o}$ is yet another constant. The conserved current
turns out to be,
\begin{equation}
Q=a\sqrt{n\phi ^{2}-2}\left( \frac{\dot{a}}{a}+\frac{n\phi
 ^{2}-1}{n\phi
^{2}-2 }\frac{\dot{\phi}}{\phi }\right)
\end{equation}
It can be verified that such a conserved current follows from
 Eqn.~(\ref
{eq:current}), upon substituting the values of $f$, $w$ and $V$. The
 form of
$w$ we have obtained in this process has an excellent feature.
 Initially
when $\phi $ is large, as should be the case, $w$ is small, finally
 when $%
\phi $ falls off, $w$ becomes large enough  leading to ordinary 
 G.R.T.

Let us now perform the change of variables to obtain the corresponding

cyclic coordinates. We need a particular solution of Eqn.
 (\ref{eq:trans}).
We find
\begin{equation}
u=\frac{a\phi ^{2}}{2}\sqrt{n\phi ^{2}-2}\quad ;\quad v=a\phi ,
\end{equation}
which can be inverted by
\begin{equation}
a^{2}=\frac{nv^{4}-4u^{2}}{2v^{2}}\quad ;\quad \phi
 ^{2}=\frac{2v^{4}}{%
nv^{4}-4u^{2}}.
\end{equation}

Being always $a > 0$, the Jacobian of transformation does not give any

problems, and the same will be for all the cases below. Under this
transformation, the Lagrangian takes the nice form
\begin{equation}
L= \frac{3 \dot{u}^2}{v} - 3 n v \dot{v}^2 + 6 k v - V_0 v^3.
\end{equation}

As announced, $u$ is cyclic. The conserved current gives
\begin{equation}
Q=\frac{\partial L}{\partial \dot{u}} = \frac{6 \dot{u}}{v}.
\end{equation}

We use now the condition $E_{L}=0$ to find $v$
\begin{equation}
(\frac{Q^2}{12}-6k)+V_{0}v^{2}=3n\dot{v}^{2},
\end{equation}
which can be integrated. Setting $F=6k-Q^{2}/12$, we get
\begin{equation}
v=\frac{-e^{\sqrt{\frac{V_{0}}{3n}}\ t}+4Fe^{-\sqrt{\frac{V_{0}}{3n}}\
 t}}{4%
\sqrt{V_{0}}},
\end{equation}
and
\begin{equation}
u=\frac{\sqrt{3}Q}{24V_{0}}\left( e^{\sqrt{\frac{V_{0}}{3n}}\
t}4Fe^{-\sqrt{%
\frac{V_{0}}{3n}}\ t}\right) +u_{0}.
\end{equation}

Here and below we set to zero the integration constant for the origin
of time. The condition $E_{L}=0$ fixes another one, so that we are
left with two ($A$ and $u_{0}$). The expression of $a$ and $\phi $ is
rather involved and we do not write it explicitly. It is interesting
anyway to show the behavior at large times. It is
\begin{equation}
a(t\rightarrow \infty )\propto e^{\sqrt{V_{0}}t}\quad ;\quad \phi
(\rightarrow \infty )=const.
\end{equation}
We see that we have inflationary asymptotic behavior and that $\phi $
 (and
then $w$) goes to a constant.

Let us observe finally that, as said before, it would be possible,
 with a
transformation of $\phi $, to bring this case to the above one. Being
 $n\neq
0$ we arrive to the nondegenerate case, so that $f$ is not in closed
 form.
Actually the transformation itself is not obtained in closed form. We
 see
thus that, although the two situations are mathematically equivalent,
 we
would pass from a solvable and physically significant situation to a
 totally
unmanageable one,  from both points of view.\medskip

{\Large \textbf{Case 3}}. \medskip

Let us consider the induced theory of gravity by choosing $f=\epsilon
 \phi
^{2}$, $\epsilon $ being the coupling constant. Under this choice,
 Eqn. (\ref
{eq:netcond}) gives
\begin{equation}
V=V_{0}\phi ^{6}\quad;\quad w=\frac{12\epsilon \phi }{n\epsilon
 ^{2}\phi
^{4}-2}
\end{equation}
along with
\begin{equation}
B_{1}=\frac{q\sqrt{n\epsilon ^{2}\phi ^{4}-2}}{\epsilon \sqrt{n}\phi
 ^{2}}
\quad;\quad B_{2}=\frac{cq\sqrt{n\epsilon ^{2}\phi ^{4}-2}}{2\epsilon
 \sqrt{n}\phi
},
\end{equation}
where, $c,q$ are constants. As a result we find
\begin{equation}
\alpha =-\frac{N\sqrt{n\epsilon ^{2}\phi ^{4}-2}}{a\phi
 ^{2}}\quad;\quad\beta =
\frac{N\sqrt{n\epsilon ^{2}\phi ^{4}-2}}{2a^{2}\phi },
\end{equation}
where, $N=\displaystyle\frac{c~l~q}{\epsilon \sqrt{n}}$ is a constant,

which we can set to unity. The conserved current turns out to be
\begin{equation}
Q=a\sqrt{n\epsilon ^{2}\phi ^{4}-2}\left(
\frac{\dot{a}}{a}+2\frac{\dot{\phi}\left(n\epsilon ^{2}\phi
 ^{4}-1\right)}{\phi\left(n\epsilon ^{2}\phi ^{4}-2\right)}\right) .
\end{equation}

Here again we observe that the conserved current follows from equation
 (\ref
{eq:current}), upon substituting the forms of $f, w$ and $V$ in it.
 Further
the form of $w$ here also has got the same excellent feature as in the

previous case.

We find now the transformed Lagrangian. As the procedure strictly
 follows
the above one, we make here and below some shortening. The
 transformation is
\begin{equation}
a^{2}=\frac{\epsilon ^{2}nv^{8}-4u^{2}}{2v^{2}}\quad ;\quad \phi
 ^{2}=\frac{%
\sqrt{2}v^{4}}{\sqrt{\epsilon ^{2}nv^{8}-4u^{2}}},
\end{equation}
and the transformed Lagrangian

 \begin{equation}
L=\frac{3\epsilon \dot{u}^{2}}{v^{2}}-12\epsilon ^{3}nv^{4}\dot{v}%
^{2}+6\epsilon kv^{2}-V_{0}v^{6}.
\end{equation}
Let us set $F=Q^{2}-72\epsilon
^{2}k$, $G=12\epsilon V_{0}$, $H=12\epsilon ^{2}\sqrt{n}$. We get
\begin{equation}
v=\frac{\sqrt{e^{\frac{2\sqrt{G}t}{H}}-4Fe^{-\frac{2\sqrt{G}t}{H}}}}{2G^{1/4}%
}\quad ;\quad u=\frac{Q}{48\epsilon G}\left(
e^{\frac{2\sqrt{G}t}{H}}+4Fe^{-%
\frac{2\sqrt{G}t}{H}}\right) .
\end{equation}

Again we obtain, for large $t$, that $a$ approaches to an exponential and $%
\phi $ to a constant.\medskip

{\Large \textbf{Case 4}}.

\medskip Let us now consider the theory of a scalar field being nonminimally
coupled with gravity, by choosing $f=1-\zeta \phi ^{2}$. As a result, we
get,
\begin{equation}
V=V_{0}(1-\zeta \phi ^{2})^{3},~~~w=\frac{12\zeta ^{2}\phi ^{3}}{(1-\zeta
\phi ^{2})[n(1-\zeta \phi ^{2})^{2}-2]},  \label{eq:cond5}
\end{equation}
and
\begin{equation}
B_{1}=\frac{\sqrt{n(1-\zeta \phi ^{2})^{2}-2}}{\sqrt{2}(1-\zeta \phi ^{2})}%
,~~~B_{2}=-\frac{c}{2\sqrt{2}\zeta }\frac{\sqrt{n(1-\zeta \phi ^{2})^{2}-2}}{%
\phi }.
\end{equation}

Hence we obtain
\begin{equation}
\alpha=-\frac{cl}{\sqrt{2}}\frac{\sqrt{n(1-\zeta\phi^2)^2-2}}{%
a(1-\zeta\phi^2)},~~~\beta=-\frac{cl}{2\sqrt{2}\zeta}\frac{\sqrt{%
n(1-\zeta\phi^2)^2-2}}{a^2\phi}
\end{equation}

Finally, we obtain the conserved  current as,
\begin{equation}
Q=\sqrt{n(1-\zeta \phi ^{2})^{2}-2}\left( \dot{a}-2\zeta \frac{n(1-\zeta
\phi
^{2})^{2}-1}{(1-\zeta \phi ^{2})(n(1-\zeta \phi ^{2})^{2}-2)}a\phi \dot{\phi}%
\right) .  \label{eq:current4}
\end{equation}

A little algebraic calculations shows that here also one can generate the
above conserved current  simply from the first integral of equation (\ref
{eq:current}), upon substituting $f,f^{\prime }f^{\prime \prime
},w,w^{\prime }$ and $V$ from solutions (\ref{eq:cond5}).

Again, we can find a transformation
\begin{equation}
a^{2}=\frac{ne^{2v}-4u^{2}e^{-2v}}{2}\quad ;\quad \phi ^{2}=\frac{1}{\zeta }%
\left( 1-\sqrt{\frac{2}{ne^{2v}-4u^{2}e^{-2v}}}\right) ,
\end{equation}
with a new Lagrangian
\begin{equation}
L=3e^{-v}\dot{u}^{2}-3e^{3v}\dot{v}^{2}+6kv-V_{0}e^{3v},
\end{equation}
and solutions
\begin{equation}
v=\log (e^{2\lambda t}-4F)-\lambda t\quad ;\quad
u=-\frac{Q\sqrt{3n}}{4V_{0}} (e^{\lambda t}+4Fe^{-\lambda t})+u_{0}
\end{equation}
where $B$ is the conserved current, $u_{0}$ is an integration constant and $%
F=3B^{2}-6k$, $\lambda =\sqrt{V_{0}/3n}$.

Asymptotic behaviors of $a$ and $\phi $ are the same as
before.\medskip

{\Large \textbf{Case 5}} \medskip

Let $\frac{w n}{\phi }f^{3}$=costant=q. Upon imposing this assumption,
following results emerge in view of equation (\ref{eq:netcond}).
\begin{equation}
\label{eq:cas5}
f=m\sqrt{1+\epsilon \phi ^{2}},~~~V=V_{0}(1+\epsilon \phi
^{2})^{\frac{3}{2}},~~~w=\frac{3}{2}\epsilon m\frac{\phi }{(1+\epsilon
\phi ^{2})^{\frac{3}{2}} },
\end{equation}
where, $m=\sqrt{\frac{2}{n}}$ and $\epsilon =\frac{qn}{6}$. $B_{1}$ and $
B_{2}$ are given by,
\begin{equation}
B_{1}=\frac{\phi }{\sqrt{1+\epsilon \phi ^{2}}},~~~B_{2}=\frac{c}{\epsilon}
\sqrt{1+\epsilon \phi ^{2}}.
\end{equation}

Hence, $\alpha$ and $\beta$ take the following form,
\begin{equation}
\alpha=-\frac{c~l}{a}(\frac{\phi}{\sqrt{1+\epsilon\phi^2}}),~~~\beta= \frac{%
c~l}{\epsilon a^2}\sqrt{1+\epsilon\phi^2}.
\end{equation}

Finally, the conserved current turns out to be,
\begin{equation}
Q=a\phi\left(\frac{\dot{a}}{a}+\frac{2\epsilon\phi^2+1}{2(1+\epsilon\phi^2)}%
\frac{\dot\phi}{\phi}\right).
\end{equation}
As before it is quite trivial to show that the first integral of equation (
\ref{eq:current}), upon substituting  f, $f^{\prime}$, $f^{\prime\prime}$,
V, w, $w^{\prime}$ form solution (\ref{eq:cas5}) in it, leads to the above
conserved current .

The transformation is
\begin{equation}
a^{2}=e^{2\epsilon v}-\frac{u^{2}}{v}e^{-2\epsilon v}\quad ;\quad \phi ^{2}=%
\frac{u^{2}}{\epsilon ^{2}\left( e^{4\epsilon v}-\displaystyle\frac{u^{2}}{%
\epsilon }\right) },
\end{equation}
so that the Lagrangian is
\begin{equation}
L=\frac{3m}{2\epsilon }e^{-\epsilon v}{\dot{u}}^{2}-6m\epsilon
^{2}e^{3\epsilon v}{\dot{v}}^{2}+6kme^{\epsilon v}-V_{0}e^{3\epsilon v},
\end{equation}
with solutions
\begin{equation}
v=\frac{1}{\epsilon }\log (e^{2\lambda t}-4F)\quad ;\quad u=\frac{HQ}{3m%
\sqrt{G}}(e^{\lambda t}+4Fe^{-\lambda t})+u_{0},
\end{equation}
where  $u_{0}$ is an integration constant and $ F=\epsilon Q^{2}-36km^{2}$,
$G=6mV_{0}$, $H=6\epsilon m$.\medskip

{\Large \textbf{Case 6}} \medskip

Let, $f=constant=f_{0}>0$ (in order to have positive Newton constant.)\\
Under this situation Eqns. (\ref{eq:i}-\ref{eq:iv}) can be solved only for
the vanishing curvature constant, i.e., for $k=0$. The method of separation
of variables yields,
\begin{equation}
A_{1}=\frac{c_{1}}{\sqrt{a}}\quad;\quad
A_{2}=-\frac{2c_{2}}{3a^{\frac{3}{2}}},
\end{equation}
where $c_{1}$ and $c_{2}$ are constants. Further, following differential
equations are obtained, viz.
\begin{equation}
\frac{V^{\prime }}{V}=\frac{9c_{1}B_{1}}{2c_{2}B_{2}}\quad;\quad\frac{B_{1}^{\prime
}}{B_{2}}=\frac{c_{2}w}{2fc_{1}\phi }\quad;\quad B_{2}-2\phi B_{2}^{\prime
}-\frac{ w^{\prime }}{w}\phi B_{2}+\frac{9c_{1}}{2c_{2}}B_{1}\phi =0.
\end{equation}
One equation is lost in the process and we have thus to impose yet another
assumption.

\textbf{Subcase 1}:

 Let $V = m^2 \phi^2$.

 Under this assumption we get
\begin{equation}
w=\frac{8f_0 \phi}{3 \phi^2 +n}
\end{equation}
It is better to treat separately the cases $n=0$ and $n\neq 0$.

\textbf{Subcase 1a : $n=0$}

In this case we have
\begin{equation}
\alpha= \frac{\phi}{\sqrt{a}}\quad;\quad \beta=-\frac{3\phi^2}{2 a^{3/2}},
\end{equation}
and the transformation is
\begin{equation}
a^3 = 3 u v \quad ; \quad \phi^2= \frac{v}{3u}.
\end{equation}

The transformed Lagrangian is
\begin{equation}
L=m^2 v^2 - 8 f_0 \dot{u}\dot{v}.
\end{equation}

The solutions are very simple and we can give directly $a$ and $\phi$
\begin{equation}
a = a_0 t^{4/3} \quad ; \quad \phi = \frac{2\sqrt{2}}{m t}.
\end{equation}

\textbf{Subcase 1b: $n\neq 0$}

We get now
\begin{equation}
\alpha =\sqrt{\frac{n+3\phi ^{2}}{a}}\quad ;\quad \beta =-\frac{3\phi \sqrt{%
n+3\phi ^{2}}}{2a^{3/2}},
\end{equation}
with transformation
\begin{equation}
a^{3}=\frac{9n^{2}u^{2}-12v^{2}}{4n}\quad ;\quad \phi
^{2}=\frac{4nv^{2}}{ 9n^{2}u^{2}-12v^{2}},
\end{equation}
and new Lagrangian
\begin{equation}
L=6f_{0}n{\dot{u}}^{2}-\frac{8f_{0}}{n}{\dot{v}}^{2}+m^{2}v^{2},
\end{equation}
with solutions
\begin{equation}
a^{3}=a_{0}\left( t^{2}-\frac{8f_{0}}{nm^{2}}\sin ^{2}\omega t\right)
\quad
;\quad \phi ^{2}=\frac{8f_{0}}{3m^{2}n}\frac{\sin ^{2}(\omega 
t)}{t^{2}-\frac{
8f_{0}}{m^{2}n}\sin ^{2}\omega t},
\end{equation}
where $a_{0}$ is an integration constant and $\omega
=\displaystyle\frac{8f_{0}}{m^{2}n}$.
Tis case differs from the others, as the asymptotic behaviour of
$a(t)$ for $t\rightarrow \infty$ is not inflationary; in fact we have
that  $a(t\rightarrow \infty)\propto t^{2/3}$, and $\phi(t\rightarrow
\infty)\propto \displaystyle\frac{\sin\left(\omega t\right)}{t}$.
\par
\textbf{Subcase 2:}\\
 $V=V_{0}\left( A\exp {(\frac{\lambda \phi }{2})}+B\exp {(\frac{-\lambda \phi }{2})}\right) ^{2}$ , with $\lambda =\frac{1%
}{2M}\sqrt{\frac{3}{2}}$, and $M^{2}=(8\pi G)^{-1}$.

This case has been extensively treated in (\cite{c:r}, \cite{g:p}, \cite
{a:p}), so that we  refer to them for the details.

\textbf{Subcase 3:}  $V=V_{0}$
This case has been treated  in  (\cite{r:sec}) and admit as conserved
current:
\begin{equation}\label{eq:conV}
  Q=a^3\phi.
\end{equation}

\section{Existence of other symmetries}

In the previous section it has been observed that for the existence of
Noether symmetry of the action (\ref{eq:action2}) in the Robertson-Walker
space time  for  $k={{\pm} }1$, $V$ is proportional to $ f^{3}$.  In this
section we shall explore other possible symmetries of the theory, which
could not be obtained by the application of Noether theorem. For this
purpose we call upon equation (\ref{eq:newform}) and note that for $ V(\phi
)$ proportional to $f^{2}(\phi )$, i.e.
\begin{equation}\label{eq:nonnet1}
V(\phi )=V_0f^{2}(\phi )
\end{equation}
where, $V_0$ is a constant, we obtain two possibilities. The first one is,
\begin{equation}
Q=(3f^{\prime 2}+\frac{2w}{\phi }f)^{\frac{1}{2}}a^{3}\dot{\phi}=constant.
\label{eq:nonnet}
\end{equation}
This is a very interesting result. It shows that symmetry exists even for
$V$ proportional to $f^{2}$, and holds for arbitrary $f$ (hence $V$ and
$\phi $), which could not be explored by previous treatment. This
interesting result was pointed out in a couple of recent publications
\cite{a:c} and \cite{a:p}%
. This conserved charge exists even for $V=0$.

There of course exists yet another possibility, viz.
\begin{equation}
3f^{\prime }{}^{2}+\frac{2wf}{\phi }=0
\end{equation}
It has been pointed out right at the beginning  that, under this condition,
the Hessian determinant vanishes; as a result the Lagrangian becomes
degenerate. Further, this equation is satisfied at the cost of making
either $f$ (and hence Newton's gravitational constant) negative, which is
of course unphysical, or the Brans-Dicke coupling parameter $w$ negative,
which leads to all time acceleration of the Universe. Further, in view of
equations (\ref{eq:nonnet1}) and (\ref{eq:nonnet}) we find that equation
(8) leads to,
\begin{equation}
\frac{d}{dt}\left( 2\dot{a}+\frac{f^{\prime }}{f}a\dot{\phi}\right) =\frac{1
}{3}V_0af.
\end{equation}
This means that there exists no conserved current in general in such a
situation other than for $V=0$.

In Sec. 3 we have underlined that, due to the choice~(18) there is the
possibility  that conditions (19) do not cover all the possible Noether
symmetries. Therefore we briefly show that the symmetry defined in
Eqn.~(\ref {eq:nonnet}) is indeed not of Noether type. To this goal we
rewrite the current~(\ref{eq:nonnet}) in the following way:
\begin{equation}
Q=G(\phi )a^{3}\dot{\phi},  \label{eq:nonnet2}
\end{equation}
where  $G(\phi )=\sqrt{3(f^{^{\prime }})^{2}+\displaystyle\frac{2wf}{\phi
}}$. If Eqn.~(\ref{eq:nonnet2}) is a Noether current, then there exists a
vector field $X$ (the generator) on $TQ$
\[
X=\alpha (a,\phi )\frac{\partial }{\partial a}+\beta (a,\phi )\frac{\partial
}{\partial \phi }+\dot{\alpha}(a,\phi )\frac{\partial }{\partial \dot{a}}+%
\dot{\beta}(a,\phi )\frac{\partial }{\partial \dot{\phi}},
\]
such that $i_{X}\theta _{L}=Q$.Thus
\begin{equation} i_{X}\theta _{L}=\left(
-6a^{2}f^{^{\prime }}\alpha +2\beta \displaystyle
\frac{w}{\phi }a^{3}-12fa\alpha \dot{a})\right) \equiv G(\phi )a^{3}\dot{\phi%
},  \label{eq:nonnet3}
\end{equation}
which implies $\alpha =0$ and $\beta =\displaystyle\frac{\phi }{2w}G(\phi )$%
. Substituting into Eqns~(14-17) gives inconstistency, unless $f=constant$.
Only in this case we have a Noether symmetry with $\alpha =0$ and $\beta
=%
\displaystyle\frac{\phi }{2w}G(\phi )$.

Due to the fact that this symmetry is generally  not of Noether type, we
cannot find cyclic variables and the treatment is more difficult. We
discuss an interesting general feature of the solution and some very
particular cases. From the expression of the conserved current, it is
possible to extract $\dot{\phi}$ and insert it into $E_{L}$. Now from
$E_{L}=0$ we obtain $\dot{a}$. If we derive $E_{L}$ w.r.t. time and use
again the expression of $\dot{a}$, we obtain an interesting expression for
$%
\ddot{a}$
\begin{equation}
\ddot{a}=\frac{2w(na^{6}f^{3}+Q^{2})+3na^{6}\phi f^{2}(f^{\prime })^{2})}{%
6fa^{5}(2fw+3\phi (f)^{\prime }{}^{2})}>0,  \label{eq:ddot}
\end{equation}
which shows that all the evolution is inflationary. In particular, when $f$
is constant, we get the very simple form
\begin{equation}
\ddot{a}=\frac{f_{0}na}{6}+\frac{N^{2}}{6f_{0}^{2}a^{5}},
\label{eq:nonnetcos}
\end{equation}
which shows that there are two regimes,  with a late time  exponential
behavior. As we said this case is indeed a special Noether symmetry.
In order to treat it like the ones above, however, we should  specify
the function $w$. If $\displaystyle\frac{w}{\phi }=\frac{1}{2}$, we
obtain a minimally coupled field with constant potential and the
conserved current is just $a^{3}\dot{\phi}$. As said above, this case
was treated in~(\cite{r:sec}). Other choices of do not to seem to be
of particular physical interest. It is in any case interesting that
Eqn.~(\ref{eq:nonnetcos}) holds independently of $w$.

In some special  cases it is possible  to use the current~(\ref{eq:nonnet})
to find particular solutions of the Einstein equations. As an example we
can consider the case $f=\phi $, $w=\displaystyle\frac{\phi }{2}$. Thus the
Eqn.~(\ref{eq:nonnet}) becomes
\begin{equation}
Q=\sqrt{3+\phi }a^{3}\dot{\phi}.  \label{eq:nonnetphi}
\end{equation}
We can obtain a solution imposing that $\phi
=\displaystyle\frac{\phi_0}{a^{2}}$, so that
\begin{equation}
\dot{a}=\frac{-|Q|}{2}\frac{a}{\sqrt{3a^{2}+1}},  \label{eq:nonnetphi2}
\end{equation}
where $G=-\left| N\right| $.With this choice the Eqn.~(\ref{eq:ddot})
becomes:
\begin{equation}
\ddot{a}=\frac{(n\phi _{0}+Q^{2})+3na^{2}\phi _{0}^{2}}{a(3a^{2}+1)}.
\label{eq:nonnetphi3}
\end{equation}
In order that the Eqs.~(\ref{eq:nonnetphi2}) and (\ref{eq:nonnetphi3}) are
compatible the following condition is required:
$|Q|=-\displaystyle\frac{2(6k-n\phi
_{0})\phi _{0}}{\sqrt{2\phi _{0}}}$. Then Eqn.~(\ref{eq:nonnetphi2})can be
solved and inverted in terms of elliptical integrals: in the following we
show the asymptotic behaviors
\begin{eqnarray}
&&a(t\rightarrow 0)\propto t  \label{eq:nonnetphi4} \\
&&\phi (t\rightarrow 0)\propto t^{-2}
\end{eqnarray}
and
\begin{eqnarray}
&&a(t\rightarrow \infty )\propto t  \label{eq:nonnetphi5} \\
&&\phi (t\rightarrow \infty )\propto t^{-2}.
\end{eqnarray}

\section{\textbf{Corresponding results in Kantowski-Sachs metric}}

We start with the same action (1). The Ricci scalar is now
$^{4}R=2(\frac{\ddot{a%
}}{a}+2\frac{\ddot{b}}{b}+2\frac{\dot{a}\dot{b}}{ab}+\frac{\dot{b}^{2}}{b^{2}%
}+\frac{1}{b^{2}})$. Hence the action, apart from a total derivative term,
is
\begin{equation}
A=4\pi \int ~[-4f^{\prime }ab\dot{b}\dot{\phi}-2f^{\prime }b^{2}\dot{a}\dot{%
\phi}-4fb\dot{a}\dot{b}-2fa\dot{b}^{2}+ab^{2}\frac{w}{\phi }\dot{\phi}%
^{2}+2fa-ab^{2}V(\phi )]dt.
\end{equation}
The Hessian determinant is $-32^{2}\pi ^{3}fab^{4}(3f^{\prime }{}^{2}+2f%
\frac{w}{\phi })$. So the Lagrangian is degenerate under the condition $%
3f^{\prime }{}^{2}+2f\frac{w}{\phi }=0$. The field equations are,
\begin{equation}
2\frac{\ddot{b}}{b}+\frac{f^{\prime }}{f}\ddot{\phi}+\frac{f^{\prime \prime }%
}{f}\dot{\phi}^{2}+2\frac{f^{\prime }\dot{b}}{fb}\dot{\phi}+\frac{\dot{b}^{2}%
}{b^{2}}+\frac{w\dot{\phi}^{2}}{2f\phi }+\frac{1}{b^{2}}-\frac{V(\phi )}{2f}%
=0
\end{equation}
\begin{equation}
\frac{\ddot{a}}{a}+\frac{\ddot{b}}{b}+\frac{f^{\prime }}{f}\ddot{\phi}+\frac{%
\dot{a}\dot{b}}{ab}+\frac{\dot{a}f^{\prime }}{af}\dot{\phi}+\frac{\dot{b}%
f^{\prime }}{bf}\dot{\phi}+\left( \frac{f^{\prime \prime }}{f}+\frac{w}{%
2f\phi }\right) \dot{\phi}^{2}-\frac{V(\phi )}{2f}=0
\end{equation}
\begin{equation}\label{eq:nonnet3}
\frac{\ddot{a}}{a}+2\frac{\ddot{b}}{b}-\frac{w\ddot{\phi}}{\phi }+2\frac{%
\dot{a}\dot{b}}{ab}+\frac{\dot{b}^{2}}{b^{2}}+\frac{(w-w^{\prime }\phi )\dot{%
\phi}^{2}}{2f^{\prime }\phi ^{2}}-2\frac{w\dot{b}\dot{\phi}}{f^{\prime
}b\phi }-\frac{w\dot{a}\dot{\phi}}{f^{\prime }a\phi }+\frac{1}{b^{2}}-\frac{%
V^{\prime }}{2f^{\prime }}=0
\end{equation}
\begin{equation}
\frac{\dot{b}^{2}}{b^{2}}+2\frac{\dot{a}\dot{b}}{ab}+\frac{f^{\prime }\dot{a}%
}{fa}\dot{\phi}+2\frac{f^{\prime }\dot{b}}{fb}\dot{\phi}-\frac{w}{2f\phi }%
\dot{\phi}^{2}+\frac{1}{b^{2}}-\frac{V(\phi )}{2f}=0
\end{equation}
In view of the above field equations one can construct yet another important
equation, viz.,
\begin{equation}
\sqrt{(3f^{\prime }{}^{2}+2\frac{wf}{\phi })}\frac{d}{dt}\left( \sqrt{%
(3f^{\prime }{}^{2}+2\frac{wf}{\phi })}ab^{2}\dot{\phi}\right) +ab^{2}f^{3}(%
\frac{V}{f^{2}})^{\prime }=0
\end{equation}
Now the detailed calculation reveals that Noether symmetry in this
situation exists at the price of making the Lagrangian degenerate. There is
nothing wrong for a Lagrangian to be degenerate. We have shown earlier
\cite{a:c}, \cite{a:p} how to deal with such Lagrangians. However, this
degeneracy indicates that either of $f$ or $w$ has to be negative, which is
nasty. It thus appears that Noether theorem does not reveals any physically
reasonable form of the coupling parameters. Nevertheless it can be shown as
before that there exists other conserved currents which can be explored
from the field equations but not from the consideration of Noether theorem.

In equation (\ref{eq:nonnet3}) if one considers $V(\phi )$ proportional to
$f(\phi )^{2}$ then last term vanishes. Hence one can choose either
$\sqrt{(3f^{\prime }{}^{2}+2\frac{ wf}{\phi })}=0$, which is the outcome of
Noether symmetry, or
\begin{equation}
Q=ab^{2}\dot{\phi}\sqrt{3f^{\prime }{}^{2}+2\frac{wf}{\phi }}=conserved.
\end{equation}
This has the same form of the conserved  current (66).
\section{\textbf{Concluding remarks}}
Once again, the Noether symmetry approach revealed a powerful tool  in the
study of scalar tensor theories.

 We have studied the more general possible action in the case of a pure scalar
 field domination.

 The first result is that asymptotic inflationary behaviour is always obtained.
 A second important result is the possibility of symmetries of more general type; a
 feature which is, to our knowledge, completely inexplored in this kind of problems.
 There is of course room for deeper investigation on this point.

 A third improvement lies in the consideration that the action~(1), more general of the
 one trated in~\cite{r:sec}, gives indeed new possibilities, despite the mathematical equivalence of the two cases.
 \\Some open problems were also discussed: there is the  clearly need of more investigation on degenerate
 cases; also the treatment in the presence of perfect fluid is important in view of
 applications to recent observations (for a recent treatment in the minimal coupling case
 see~\cite{ru:s},~\cite{c:al}.
 But most of all a good understanding of the physical meaning of Noether
 symmetries in this context would be the greatest  hit.
\section{Appendix
}
 In this section we want at least sketch the calculation, leading to the
Eqs.~(\ref{eq:netcond}) starting from the $\mathit{factoring}$ hypothesis
in Eqn.~(\ref{eq:fact}). In order to perform this let us start from the
Eqn.~( \ref{eq:i}) and divide by $B_{1}$, obtaining:
\begin{equation}
\frac{A_{1}+2aA_{1,a}B_{1}}{a(A_{2}+2aA_{2,a})}=-\frac{B_{2}}{B_{1}}\frac{
f^{^{\prime }}}{f}=-C_{1}, \label{eq:ap1}
\end{equation}
where we assume $(A_{2}+2aA_{2,a})\neq 0$. The case $(A_{2}+2aA_{2,a})=0$
will be discussed later. From the Eqn.~(\ref{eq:ii}) we obtain
\begin{equation}
a\frac{A_{2}}{A_{1}}=3\phi \left( \frac{B_{1}-2\phi \frac{f^{^{\prime }}}{w}%
B_{1}^{^{\prime }}}{B_{2}-2\phi B_{2}^{^{\prime }}-\phi \frac{w^{^{\prime }}%
}{w}B_{2}}\right) =C_{2}  \label{eq:ap2}
\end{equation}
where we assume that $B_{2}-2\phi B_{2}^{^{\prime }}-\phi \frac{w^{^{\prime
}}}{w}B_{2}\neq 0$. Using the Eqs.~(\ref{eq:ap1}) and (\ref{eq:ap2}), we
can replace $A_{2}$ and $B_{2}$ in Eqn.~(\ref{eq:iv}), obtaining the
following relation:
\begin{equation}
6kf\left( 1+C_{1}C_{2}\right) =a^{2}V\left( 3+C_{1}C_{2}\frac{V^{^{\prime
}} }{V}\frac{f}{f^{^{\prime }}}\right) ,  \label{eq:ap3}
\end{equation}
which, for $k\neq 0$ implies that:
\begin{eqnarray}
C_{1}C_{2} &=&-1  \label{eq:ap4} \\
V &=&V_{0}f^{3}.
\end{eqnarray}
In order to obtain $A_{1}$ and $A_{2}$, let us set
$C_{1}=c=-\frac{1}{C_{2}}$ , and replace in the Eqn.~(\ref{eq:ap1}) and
(\ref{eq:ap2}), so that
\begin{eqnarray}
 \label{eq:app}
A_{2} &=&\frac{l}{a^{2}}  \\ A_{1} &=&-\frac{cl}{a}. \\ B_{2}
&=&c\frac{f}{f^{^{\prime }}}B_{1}
\end{eqnarray}
Finally, using the Eqn.~(\ref{eq:iii}) and~(\ref{eq:ap2}) we obtain the
last two equations in ~(\ref{eq:netcond}):
\begin{eqnarray}
 \label{eq:ap6}
&&B^{^{\prime}}_1=\frac{2}{3}\frac{w}{\phi f^{^{\prime}}}B_1, \\
&& 3{f^{^{\prime}}}^2+2\frac{w}{\phi}f=n\frac{wf^3}{\phi}.
\end{eqnarray}
We note that the Eqns~(\ref{eq:app}--95) imply the following general
relation between  $\alpha$ and $\beta$:
\begin{equation}\label{eq:albet}
\alpha=-a\beta\frac{f^{'}}{f}.
\end{equation}
Since the Eqn~(\ref{eq:iv}) this implies that both the sides in the
\ref{eq:iv} vanishes  separately , as was supposed  {\it  a  priori}
in~\cite{c:r}. Let us go back to dismiss the assumptions put forward above
in the calculations, that is $(A_2+2aA_{2,a})\neq 0$ and $B_2-2\phi
B^{^{\prime}}_2-\phi\frac{ w^{^{\prime}}}{w}B_2 \neq 0$. If
$(A_{2}+2aA_{2,a})=0$, then also $(A_{1}+2aA_{1,a})=0$, so that
\begin{eqnarray}
A_{1} &=&\frac{q}{\sqrt{a}}  \label{eq:ap7} \\
A_{2} &=&\frac{p}{a},
\end{eqnarray}
while $B_{2}\frac{f^{^{\prime }}}{f}$ and $B_{2}$ remain arbitrary (but $%
\neq 0$). However such solutions are incompatible with the Eqn.~(\ref{eq:ii}%
) unless $a=0$, which is a unphysical trivial solution. Consider now the
case $B_{2}-2\phi B_{2}^{^{\prime }}-\phi \frac{w^{^{\prime }}}{w}B_{2}=0$,
which implies that $B_{1}-2\phi \frac{f^{^{\prime }}}{w} B_{1}^{^{\prime
}}=0$, i.e.
\begin{eqnarray}
\frac{B_{1}^{^{\prime }}}{B_{1}} &=&\frac{w}{2\phi f^{^{\prime }}}
\label{eq:ap8} \\
B_{2}^{2} &=&\frac{b\phi }{w}.
\end{eqnarray}
Such equations are incompatible with Eqn.~(\ref{eq:iii}), unless one reduces
to the degenerate case
\[
3{f^{^{\prime }}}^{2}+2\frac{w}{\phi }f=0.\]
\section*{Acknowledgment}
This work has been carried out during the visits of A.K. Sanyal to
I.C.T.P. and to University ''Federico II'', Naples, Italy. Thanks are due
to both the Institutions and to P.R.I.N. ''SINTESI'' for their support and
hospitality.

\end{document}